\documentclass[12pt,preprint]{aastex}

\usepackage{graphicx,epstopdf}
\begin{document}

\title{Trajectory of  a light ray in Kerr field: \\
        A material  medium approach}

\author{Saswati Roy  \altaffilmark{1,2} and A. K. Sen\altaffilmark{1,3}}

\altaffiltext{1}{Physics Department, Assam University, Silchar-788011, India}
\altaffiltext{2}{sr.phy2011@yahoo.com}
\altaffiltext{3}{asokesen@yahoo.com}

\begin{abstract}
The deflection of light ray as it passes around a gravitational
mass can be  calculated by different methods. Such calculations
are generally done by using the null geodesics under both strong
field and weak field approximation. However, several authors have
studied the gravitational deflection of light ray using material
medium approach. For a static, non-rotating spherical mass, one
can determine the deflection in Schwarzschild  field, by
expressing the line element in an isotropic form and calculating
the refractive index to determine the trajectory of the light ray.
In this paper, we draw our attention to the refractive index of
light ray in Kerr field using the material medium approach. The
frame dragging effects in Kerr field was considered to calculate
the velocity of light ray and finally the refractive index in Kerr
field geometry was determined. Hence the deflection of light ray
in Kerr field  was calculated, assuming far field approximation
and compared the  results with those calculations done earlier
using Null geodesics.
\end{abstract}





\section{\label{1}Introduction}

Gravitational deflection of light ray is the most important
consequence of Einstein's General theory of Relativity.
Scientists have worked to find out the light
deflection angle due to a gravitating body using
different approaches. One of the approaches is the
null geodesics, where based on any of the forms of the
line element, either by using the perturbation \citep{Mis72,dInv98} or
by integrating the null geodesic equations \citep{Chand83,Wein72,Wald84}, the
deflection of light ray is calculated. In this paper we
have considered the \emph{material medium approach}, to find
out the trajectory of the light ray  in the gravitational
field of a rotating body (also known as Kerr field).
In this approach, the effect of the
gravitation on the ray of light is estimated by
considering the propagation of the electromagnetic waves
be reduced to the problem of wave propagation in
material medium in flat space time. This concept
can be defined as the equivalent material medium
approach. This method is attractive because it
suggests that the classical optics is as suitable
as that of the Riemannian geometry for the study
of the electromagnetic phenomena in a gravitational
field.

Material medium approach was first used by the
author Tamm \citep{Tam24} and then  by Balaz
\citep{Bal58}, to calculate the effect of a rotating body on the
polarization of light. The same concept had also been
utilized by Plebanski \citep{Pleb60} to study
the light scattering by gravitational field. Felice
\citep{Fel71} had also discussed the optical phenomena for
the deflection of a electromagnetic wave by
gravitational field. Mashhoon \citep{Mash73,Mash75} had calculated
the deflection and polarization due to the
Schwarzschild and Kerr black holes. Later Kopeikin
and Mashhoon \citep{Kop02} also followed the same approach, to correct for the light deflection angle
of the Shapiro time delay caused by the rotation
of gravitating bodies.
 Fishbach and Freeman \citep{Fis80} derived the
effective refractive index of the material medium in
Schwarzschild field and calculated the second order
contribution to  the gravitational deflection, where
the value of the refractive index of any
medium indicates the strength of the gravitational
field \citep{Land98}. In a recent series of articles,
Evans and his co-workers
\citep{Evans86,Evans96a,Evans96b,Nan95} derived and used the effective refractive index
to calculate the gravitational time delay and
trajectories of light rays in Schwarzschild geometry.
The authors also showed that the Opto-mechanical
analogy of general relativity reproduces the equation
of \textbf{GR} and matches with the classical equations.
P. M. Alsing \citep{Als98} has extended the Newtonian
formalism of Evans, Nandi and Islam \citep{Evans96b} to the case
of stationary metrics, typical of rotating space times.

Ishihara, Takahashi and Tmimatsu  \citep{Ish88} studied how
the polarization vector of a linearly polarized
electromagnetic wave propagates in a curved
space-time in the presence of Kerr black hole.
They have showed that, in the weak field limit
the rotation angle of the plane of polarization is
proportional to the line of sight component of the
black hole's angular momentum.
The above fact is very similar to the Faraday
Effect, if the angular momentum is replaced by the
magnetic field. This justified that, the rotation of
the plane of polarization due to the angular momentum
of the Kerr black hole is just the {\it Gravitational
Faraday Rotation}. Sereno \citep{Ser03,Ser04} has also used the
similar idea to derive the time delay function and
deflection angle by drawing the trajectory of the light
ray by Fermat's principle and also discussed the
\textit{Gravitational Faraday Rotation} in the weak field limit.

Very recently, Sen \citep{Sen10} also used the same approach to
calculate the light deflection for a static non rotating
mass in Schwarzschild geometry, by expressing the line
element in an isotropic form to determine the
trajectory of the light ray. In this paper
we shall follow a similar approach corresponding
to Kerr field.

On the other hand, in a more conventional
manner, using null geodesic method many authors
have worked to find out the deflection
angle solutions in the Kerr field.
Iyer and Hansen \citep{Iye09} calculated the
deflection angle using the null geodesic in the
equatorial plane. They also concluded that
the deflection angle due to Kerr field is greater
than the Schwarzschild value for pro-grade or direct
orbit and smaller for retrograde orbit. But as
in the schwarzschild case, when the deflection angle
exceeds $2\pi$, it forms in multiple loops  and
relativistic images are produced. For the higher rotation
case, the effect is much more pronounced.

Bray \citep{Bray86}  presented
approximate solutions to the equation
of motion for a ray of light in the Kerr field
and considered multi-imaging aspect of the
gravitational lens effect.
Bozza \emph{et al.}\citep{Boz05} presented an
analytical treatment of gravitational lensing by
Kerr black holes, restricting the observers in the
equatorial plane. Keeton and Petters \citep{Ket05,Ket06a,Ket06b} have
developed a general formalism for lensing by
spherically symmetric lenses and studied the
gravitational lensing by compact objects
considering the metric in Taylor series
expansion in $\frac {G M}{c^2r}$. Ye and Lin \citep{Ye08},
discussed the strong similarities between
gravitational lensing and optical lensing using the
graded refractive index approach.

In the present paper we used the line element under
Kerr field in Boyer Lindquist form \citep{Boyer67} to calculate
the gravitational bending of a light ray. In Section \ref{2},
we have calculated the velocity and hence the refractive
index of a light ray in the Kerr field . The
value of $\frac{1}{c}\frac{d\phi}{dt}$ involved in the expression of refractive
index, due to rotating mass is also evaluated. In
Section \ref{3}, we have obtained the expression of deflection
angle of a light ray in equatorial plane due to Kerr
gravitating mass. Section \ref{4} is devoted to
make the numerical calculation of the
refractive index and deflection angle due to Sun and
some other Pulsars. In Section \ref{5}, we have concluded
our work.

\section{\label{2}The refractive index  of a light ray in Kerr field}

The exact solution of Einstein's Field Equation of
General Relativity for a stationary, axially symmetric
gravitational field of an uncharged rotating body is
given by the Kerr metric. This celebrated solution
was first given by R. P. Kerr \citep{Kerr63}. In the Boyer
Lindquist form, in the $(ct,r,\theta,\phi)$ co-ordinate system,
the line element of Kerr metric is as follows:

\begin{eqnarray}\label{E1}
ds^2= &&(1-\frac{r_gr}{\Sigma^{2}})c^2dt^2 - \frac{\Sigma^{2}}{\Delta}dr^2 - \Sigma^{2}d\theta^2 - (r^2+\alpha^2+\frac{r_gr\alpha^2}{\Sigma^{2}}sin^2\theta)sin^2\theta d\phi^2\nonumber\\
&&+\frac{2r_gr\alpha}{\Sigma^{2}}csin^2\theta d\phi dt
\end{eqnarray}

where
\begin{mathletters}
\begin{equation}\label{E2a}
\Sigma^2=r^2+\alpha^2cos^2\theta
\end{equation}
and
\begin{equation}\label{E2b}
\Delta=r^2-r_g r+\alpha^2
\end{equation}
\end{mathletters}

The constants $r_g$ and $\alpha$ are the Schwarzschild
radius and the rotation parameter of the Kerr
gravitating body. As usual $r_g=\frac {2G M}{c^2}$  and
$\alpha$ is defined as   $\alpha=\frac{J}{Mc}$, where $J$
is the angular momentum of the gravitating body, M
is the total mass of the gravitating body and $c$ is
the velocity of light.

If we consider $\alpha\rightarrow0$ the
Kerr line element  reduces to Schwarzschild line element,
which represents the non-rotating or static gravitating
mass. Like the Schwarzschild line element,
the Kerr element is also  asymptotically flat.
The coefficients of line element are
independent of $\phi$ showing that it is axially symmetric.

In equatorial plane $\theta=\frac{\pi}{2}$, which implies $\Sigma^2=r^2$.
Under far field approximation we can assume,
$\frac{\alpha^{2}}{r^{2}}<<1$. Thus the linearized form of the Kerr
 metric  in terms of spherical polar co-ordinates  $r, \theta, \phi$ can be obtained
with the first power in $\frac{\alpha}{r}$ as \citep{Isl09,Vogt05}:

\begin{eqnarray}\label{E3}
ds^2=&&(1-\frac{r_g}r)c^2dt^2 -\frac {1}{(1-\frac {r_g}{r})}dr^2- r^2 d\theta ^2
- r^2(1  +\frac{\alpha^{2}}{r^{2}}+ \frac{r_g \alpha^2}{r^3} )d \phi^2
+\frac{2r_g\alpha}{r}c d\phi dt\nonumber\\
or,    ds^2\cong  &&[(1-\frac{r_g}{r}) + \frac{2r_g \frac{\alpha}{c}
 }{r} \frac{d\phi}{dt}]c^2dt^2 -  \frac{1}{(1-\frac
{r_g}{r})}dr^2 - r^2 ( d\phi^2 + d\theta^2)
\end{eqnarray}

In most practical purposes the linearized Kerr metric
satisfactorily describes the gravitational field around
a rotating star or planet. To derive the expression for
refractive index, we follow a procedure similar to the
one adopted by Sen \citep{Sen10} for a static field (Schwarzschild
geometry). Thus to express the above line element in
an isotropic form we introduce a new radius
co-ordinate ($\rho$) with the following transformation
equation \citep{Land98} as

\begin{equation}\label{E4}
\rho= \frac{1}{2} [(r-\frac{r_g}2)+ r^{1/2} (r-r_g)^{1/2}]
\end{equation}

The above equation may be also  written as:

\begin{equation}\label{E5}
r = \rho (1 + \frac {r_g} {4 \rho})^2
\end{equation}

As was done by Sen \citep{Sen10} from Eqn. (\ref{E5}) the  value of
$\frac{dr}{d\rho}$ can be calculated as

\begin{eqnarray}\label{E6}
\frac{dr}{d\rho}= && (1+ \frac{r_g}{4\rho})^2 - \frac{r_g}{2\rho}(1+\frac{r_g}{4\rho})\nonumber\\
=&& 1-\frac{r_g^2}{16\rho^2}
\end{eqnarray}

Substituting the value of $r$ and $dr^2$ from Eqn. (\ref{E5})
and (\ref{E6}) in Eqn. (\ref{E3}) (which has far field or slow rotation
approximation) we get:

\begin{eqnarray}\label{E7}
ds^2=&&[1-\frac{r_g}{\rho(1+\frac{r_g}{4\rho})^2}+\frac{2r_g\frac{\alpha}{c} }{\rho(1+\frac{r_g}{4\rho})^2}\frac{d\phi}{dt}]c^2dt^2
- [\frac{(1-\frac{r_g^2}{16\rho^2})^2}{1-\frac{r_g}{\rho(1+\frac{r_g}{4\rho})^2}}]d\rho^2
- \rho^2(1+ \frac{r_g}{4\rho})^4( d\phi^2 + d\theta^2)\nonumber\\
=&&[\frac{(1-\frac{r_g}{4\rho})^2+2\frac{r_g}{\rho}\frac{\alpha}{c} \frac{d\phi}{dt}}
{(1+\frac{r_g}{4\rho})^2}]c^2dt^2
- (1+\frac{r_g}{4\rho})^4 [d\rho^2 + \rho^2(  d\phi^2 + d\theta^2)]
\end{eqnarray}

The above result expressed by Eqn.(\ref{E7}) gives the
isotropic form of Kerr solution.

Now in spherical co-ordinate system the quantity
$(d\rho ^2 + \rho^2 (d \phi ^ 2 + d \theta ^ 2))$ has the dimension of square
of infinitesimal length vector $d \overrightarrow{\rho}$.

By setting $ds=0$,  the  velocity of light $(v(\rho,\theta))$
can be identified from the expression of the form
$ds^2 = f (\rho,\theta)dt ^2 - d\overrightarrow{ \rho} ^2 $, as $v(\rho,\theta)= \sqrt {f(\rho,\theta )}$.
Therefore the velocity of light in the present case
(characterized by Schwarzschild radius $r_g$ and rotation
parameter $\alpha$) can be expressed as:

\begin{equation}\label{E8}
v( \rho,\theta ) = \frac{\sqrt{(1-\frac{r_g}{4\rho})^2+
2\frac{r_g}{\rho}\frac{\alpha}{c} \frac{d\phi}{dt}}}{(1+\frac{r_g}{4\rho})^3}c
\end{equation}

But this expression of velocity of light is in the unit of
length $\rho$ per unit time. We therefore write

\begin{eqnarray}\label{E9}
v(r,\theta) =&& v(\rho,\theta) \frac {dr}{d\rho}\nonumber\\
=&& v(\rho,\theta)[1-\frac{r_g^2}{16\rho^2}] \nonumber\\
=&&c(\frac{4\rho-r_g}{4\rho+r_g})^2\sqrt{1+8 \frac{\alpha}{c} \frac{d\phi}{dt} \frac{4\rho r_g}{(4\rho-r_g)^2}}
\end{eqnarray}

Substituting the value of $\rho$ from Eqn. (\ref{E4}) as
$4\rho=2r- r_g + 2\sqrt{r(r-r_g)}$ and then replacing
$r/r_g$ by $x$, we can write the above expression for
velocity of light as :

\begin{eqnarray}\label{E10}
v^2(x,\theta) =&& c^2(1-\frac{1}{x})^2 (1+ 2 \frac {\alpha}{c}\frac{ 1}{x-1}  \frac {d\phi}{dt})
\end{eqnarray}

Similarly,  the refractive index $n(x,\theta)$ can be expressed by
the relation:

\begin{eqnarray}\label{E11}
n(x,\theta)=&&\frac{x}{x-1}[1+ 2 \frac{\alpha}{c} \frac{1 }{x-1}  \frac{d\phi}{dt} ]^{-\frac{1}{2}}
\end{eqnarray}

when  $\alpha=0$, we find the central gravitational
mass is static and in that case the above
expression of refractive index goes over
to that for Schwarzschild mass, which is :

\begin{eqnarray}\label{E12}
n(r)=&&\frac{x}{x-1}=\frac{r}{r-r_g}
\end{eqnarray}

This is exactly same as the refractive index calculated
by Sen\citep{Sen10} for a static non rotating mass
(Schwarzschild geometry). Thus the second term inside
the square bracket of RHS of Eqn. (\ref{E11})
is due to the rotation of the gravitating body. The second term consists
of the rotation parameter $\alpha$
and the frame-dragging parameter $\frac{d\phi}{cdt}$ due to the rotating body.
For a static body, this term just vanishes. However, for a
very slow rotating body RHS of Eqn. (\ref{E11}) can be expanded,
as an infinite converging series, as the
second term is $<<$ 1.
It may be noted here that Sen\citep{Sen10}'s work is restricted to
static geometry. But Eqn. (\ref{E11}) can include, the case of a rotating body
(stationary geometry). Therefore, the case discussed by Sen\citep{Sen10}, comes out as a
special case when $\alpha$ is set to be zero.

\subsection{\label{2.1}Calculation of $d\phi/dt$ in the expression of refractive index }

The value of $d\phi/dt$ can be calculated by following a
procedure from Landau and Lifshitz \citep{Land98}. Below we
outline this procedure, which can be used to calculate
the expression for $d\phi/dt$.

In the gravitational field of a rotating spherical
mass, the relativistic action function $S$ for a particle
with the time $t$ and the angle $\phi$ as cyclic variables,
can be expressed as:

\begin{equation}\label{E13}
S=-{{\emph{E}}_0t}+L\phi+S_r(r)+S_\theta(\theta)
\end{equation}

where $\emph{E}_0$ is the conserved energy and $L$ denotes the
component of the angular momentum along the axis
of the symmetry of the field.

 The four momentum of the particle is

\begin{equation}\label{E14}
p^{i} = mc\frac{dx^i}{ds} = g^{ik}p_k = -g^{ik}\frac{\partial{S}}{\partial{x^k}}
\end{equation}

where $i$ and $k$  have the values 0,1,2,3 which
stand for the coordinates $ct,r,\theta,\phi$ respectively \citep{Land98}.
Now, for the variables $t$ and $\phi$
one can write :

\begin{equation}\label{E15}
 mc\frac{dx^0}{ds}  = -g^{00}\frac{\partial{S}}{\partial{x^0}}-g^{01}\frac{\partial{S}}{\partial{x^1}}
 -g^{02}\frac{\partial{S}}{\partial{x^2}}-g^{03}\frac{\partial{S}}{\partial{x^3}}
\end{equation}

and
\begin{equation}\label{E16}
mc\frac{dx^3}{ds}  = -g^{30}\frac{\partial{S}}{\partial{x^0}}-g^{31}\frac{\partial{S}}{\partial{x^1}}
-g^{32}\frac{\partial{S}}{\partial{x^2}}-g^{33}\frac{\partial{S}}{\partial{x^3}}
\end{equation}

Comparing with the  Kerr line element expressed by Eqn. (\ref{E1})
one can write:

\bigskip

 $g_{00}=(1-\frac{r_g r}{\Sigma^2}),\qquad  g_{11}=-\frac{\Sigma^2}{\Delta}, \qquad g_{22}= -\Sigma^2 ,\\[5pt]
g_{33}=-(r^2 + \alpha^2 + \frac{r_g r \alpha^2}{\Sigma^2}\sin^2\theta)\sin^2\theta ,\\[5pt]
\qquad g_{03}=g_{30}=\frac{r_g r \alpha}{\Sigma^2}\sin^2\theta$
\bigskip

The determinant of the linearized form of the
Kerr metric tensor is given by \citep{Land98,Wiltshire09}:

\begin{equation}\label{E17}
|g| =- \Sigma^{4}\sin^2\theta
\end{equation}

Thus using the formula $g^{ij}=\frac{co-factor of g_{ij}}{|g|}$
the contravariant components are:

$g^{00}=\frac{1}{\Delta  }(r^2 + \alpha^2 + \frac{r_g r \alpha^2}{\Sigma^2}\sin^2\theta)$,
\bigskip

$g^{33}=-\frac{\Sigma^2 -r_g r}{\Sigma^2 \Delta \sin^2\theta}$,
\bigskip

$g^{03}=g^{30}=\frac{r_g r \alpha }{\Sigma^2 \Delta  }$
\bigskip

and other components are zero.

Using the above values of the components of the
metric tensors, the Eqn.(\ref{E15}) and (\ref{E16}) become

\begin{eqnarray}\label{E18}
mc^2\frac{dt}{ds} = -\frac{1}{\Delta  }(r^2 + \alpha^2 + \frac{r_g r \alpha^2}{\Sigma^2}\sin^2\theta)(-\frac{\emph{E}_0}{c})- \frac{r_g r \alpha }{\Sigma^2 \Delta  }L
\end{eqnarray}

\begin{equation}\label{E19}
mc\frac{d\phi}{ds} = - \frac{r_g r \alpha }{\Sigma^2 \Delta  } (-\frac{\emph{E}_0}{c})+ \frac{\Sigma^2 -r_g r}{\Sigma^2 \Delta \sin^2\theta}L
\end{equation}

Therefore, the value of $\frac{d\phi}{dt}$ is

\begin{eqnarray}\label{E20}
\frac{d\phi}{dt}=&& \frac{r_g r \alpha \sin^2\theta \frac{\emph{E}_0}{c}+ (\Sigma^2 -r_g r)L}{[\Sigma^2  (r^2 + \alpha^2 + \frac{r_g r \alpha^2}{\Sigma^2}\sin^2\theta)\frac{\emph{E}_0}{c}- r_g r \alpha L] \sin^2\theta }c
\end{eqnarray}

In the propagation of a light ray ( or photon like
particle ), the momentum $(p)$ and the conserved
energy ( $\emph{E}_0$ ) can be expressed by the relation

\begin{mathletters}
\begin{eqnarray}\label{E21a}
\emph{E}_0 = pc
\end{eqnarray}

 In the beginning of our problem, we first considered
the general three dimensional form of the Kerr solution.
However, to make the problem mathematically simpler to solve,
we later defined our geometry contained in the equatorial plane
by choosing $\theta=\frac{\pi}{2}$ (in Section \ref{2}).
By applying such conditions to the original Kerr solution
expressed by Eqn.(\ref{E1}), we get a simpler form of the Kerr
solution as Eqn.(\ref{E3}). After that we applied the material medium
approach to get the value of refractive index as expressed in Eqn.(\ref{E11}).

But in Section \ref{2.1}, to calculate the frame dragging parameter
$\frac{d\phi}{cdt}$ of Kerr gravitating body, we
have not restricted our work to the equatorial plane in the begining. So we
obtained a general expression for frame-dragging as in Eqn.(\ref{E20}).

Later, we calculate the frame-dragging by restricting our problem to the equatorial plane,
as it has been already done in Section \ref{2} to make the calculation easy.

 Thus, by restricting the light ray in the
equatorial plane $(\theta=\frac{\pi}{2})$, the angular momentum
$L$ can be expressed as

\begin{equation}\label{E21b}
L= pb
\end{equation}
\end{mathletters}

where $b$= impact parameter, which implies

\begin{equation}\label{E22}
\frac{Lc}{\emph{E}_0}=  b
\end{equation}

So, for equatorial plane, $\Sigma^2=r^2$ and the Eqn.(\ref{E20}) becomes

\begin{eqnarray}\label{E23}
\frac{d\phi}{dt}=&& \frac{r_g \alpha    +  (r  -r_g  )b}{r^3( 1+ \frac{\alpha^2}{r^2} + \frac{r_g r \alpha^2}{r^4})- r_g  \alpha b}c
\end{eqnarray}

Now under far field approximation  (corresponding to the line element
as in Eqn. (\ref{E3})), the above expression becomes

\begin{eqnarray}\label{E24}
\frac{d\phi}{dt}=&&\frac{r_g   \alpha    +  (r  -r_g  )b}{r^3  - r_g   \alpha b  }c\nonumber\\
=&&\frac{x v +  (u-v)}{r_g(x^3  -   u v)  }c
\end{eqnarray}

where we define $\frac{r}{r_g}=x$ (as was done earlier), $\frac{\alpha}{r_g}=u$
and $\frac{b}{r_g}=v$.

Now by substituting the value of $d\phi/dt$ from Eqn.(\ref{E24})
into Eqn. (\ref{E10}), the velocity of propagation of light
ray in Kerr geometry can be expressed as :

\begin{eqnarray}\label{E25}
v^2(x,\frac{\pi}{2})=&& c^2(1-\frac{ 1 }{x})^2 (1+ 2 \frac{\alpha}{c}\cdot\frac{1}{(x-1)}\cdot\frac{x v +  (u-v)}{r_g(x^3  - u v)  }c )\nonumber\\
=&& c^2(1-\frac{ 1 }{x})^2 (1+ 2 u  \frac{x v + u  -  v}{(x-1)(x^3  - u v )})\nonumber\\
=&& c^2(1-\frac{ 1 }{x})^2 + 2 u c^2(1-\frac{ 1 }{x})^2 \frac{x v + u  -  v}{(x-1)(x^3  - u v )}
\end{eqnarray}

In the above expression of velocity of light
we find, the first term refers to the velocity due
to Schwarzschild geometry alone \citep[Sen][]{Sen10} and the second
term refers to the contribution due to rotation
(under Kerr field geometry).

From Eqn.(\ref{E25}), we can write the refractive index
$n(x,\frac{\pi}{2})$ at an arbitrary  point on equatorial
plane in the Kerr field  as:

\begin{eqnarray}\label{E26}
n(x,\frac{\pi}{2})=&&  \frac{x }{x-1}  [1+ 2 u  \frac{x v + u  -  v}{(x-1)(x^3  - u v )}]^{-\frac{1}{2}}\nonumber\\
=&&   n_{0}(x)  [1+ 2 S_{x}]^{-\frac{1}{2}}\nonumber\\
=&&n_{0}(x).\eta_{x}
\end{eqnarray}

In the above, we introduced the parameters

$n_{0}(x)=\frac{x}{x-1}$ , $S_{x} =  \frac{u(x v + u  -  v)}{(x-1)(x^3  - u v )}$
and $\eta_{x}=[1+ 2 S_{x}]^{-\frac{1}{2}}$.

Here we can also show for all $r>r_g$, we get $S_{x}<<1$.
This  is possible, as when $r >> r_g$ and $r >> \alpha$ , we must
 have $x >> 1$ and $x >>u$. Also since  $\alpha < b$, we must have
 $u < v$.  Now as $x >> 1$, we can approximate $(x-1)\sim x$ and then
 we can finally show that $S_x << 1$.  Therefore, one
can also write the expression of refractive index
in terms of the following converging series :

\begin{eqnarray}\label{E27}
n(x,\frac{\pi}{2})=&&\frac{x}{x-1}(1-   S_{x} + \frac{3}{2} S_{x}^2.....)
\end{eqnarray}

Now at $\alpha=0$, $S_{x}=0$ so that the refractive index
becomes $n(x,\frac{\pi}{2})=\frac{x}{x-1}=n_{0}(x)$, which is exactly the
same result calculated by Sen\citep{Sen10}. As already
discussed in the end of section \ref{2}, $n(x,\frac{\pi}{2})$
gives the general expression for refractive index in Kerr geometry
as a converging infinite series. For a slow rotating body the series
will converge much faster. However, for static body the term containing $S_x$,
$S_x^{2}$ etc. in the RHS of Eqn.(\ref{E27}) will become zero. This will make the
refractive index equal to $\frac{x}{x-1}$, which is the expression
already obtained by Sen\citep{Sen10}.

\section{\label{3}Calculation of Deflection in Kerr field }

 The Eqn. (\ref{E27}) provides the general expression for
refractive index on the equatorial plane in Kerr field.
Thus, the trajectory of the light ray
 can be written as \citep[Sen][]{Sen10,Born59}:

\begin{equation}\label{E28}
\triangle \psi = 2 \int^{\infty}_{b} { \frac {dr}{r
\sqrt{(\frac{n(r).r}{n(b).b})^2-1}}}- \pi
\end{equation}

 As had been already discussed by Sen\citep{Sen10}, here in our
 present problem the light is approaching from
 asymptotic infinity ( $r=-\infty $ or $x=-\infty $)
 towards  the rotating gravitational mass,
 which is placed at the origin and characterized
 by Schwarzschild radius $r_g$ and rotation
 parameter $\alpha$. Then the ray goes to  $r=+\infty $,
or $x=+\infty $ after undergoing certain
amount of deflection ($\triangle \psi$).
 Here, the closest distance of approach, for the
 approaching ray is $b$. (we note that, in actual case the
 impact parameter and closet distance of approach are
 only approximately equal.)
 When the light ray passes through
 the  closest distance of approach (i.e.$r=b $), the tangent
 to the trajectory becomes perpendicular to the vector
 ${\overrightarrow r}$ ( which is $ {\overrightarrow b}$).

  Now we change the variable to $x=\frac{r}{r_g}$ so that
  $dr= r_g   dx$ . The corresponding limit changes from
  $x=v$ to $x=\infty$, as the limit of $r$ changes from
  $r=b$ to $r=\infty$.

Accordingly, the value of deflection ($\triangle \psi$), can be
written as :

\begin{eqnarray}\label{E29}
\triangle \psi =&& 2 \int^{\infty}_{v} { \frac {dx}{x\sqrt{(\frac{n(x)\cdot x}{n(v)\cdot v})^2-1}}} - \pi\nonumber\\
=&&2 I - \pi
\end{eqnarray}

Using Eqn. (\ref{E26}) and substituting $D_{k}= n(v)\cdot v$, like it
was done by Sen \citep{Sen10}, we can re-write the above
equation as :

\begin{eqnarray}\label{E30}
I=&& n(v)\cdot v \int^{\infty}_{v} { \frac {dx}{x\sqrt{(n(x)\cdot x)^2-(n(v)\cdot v)^2}}}\nonumber\\
=&& D_{k} \int^{\infty}_{v} { \frac {dx}{x\sqrt{\{n_{0}(x)  (1+ 2S_{x})^{-\frac{1}{2}}\cdot x\}^2- D_{k}^2}}}\nonumber\\
=&& D_{k} \int^{\infty}_{v} \frac{ dx}{x\sqrt{n_{0}^{2}(x) x^{2} -D_{0}^2+n_{0}^{2}(x) x^{2} (1+ 2S_{x})^{-1}- n_{0}^{2}(x) x^{2}+D_{0}^2-D_{k}^2  }}\nonumber\\
=&& D_{k} \int^{\infty}_{v} \frac{ dx}{x\sqrt{n_{0}^{2}(x) x^{2}-D_{0}^2}}[1 + \frac{n_{0}^{2}(x) x^{2} \{(1+ 2S_{x})^{-1}  -  1\} +D_{0}^2-D_{k}^2 }{n_{0}^{2}(x) x^{2}-D_{0}^2}]^{-\frac{1}{2}}\nonumber\\
=&& D_{k} \int^{\infty}_{v} \frac{ dx}{x\sqrt{n_{0}^{2}(x) x^{2}-D_{0}^2}}[1 + J(x)]^{-\frac{1}{2}}
\end{eqnarray}

where  $D_{0}=n_{0}(v).v$ (corresponding to schwarzschild
deflection). And we have also denoted

\begin{eqnarray}\label{E31}
J(x)=\frac{n_{0}^{2}(x) x^{2} \{(1+ 2S_{x})^{-1}  -  1\}+D_{0}^2-D_{k}^2 }{n_{0}^{2}(x) x^{2}-D_{0}^2}
\end{eqnarray}

At this stage we can show that $ J(x)<< 1$.
As $J(x)$ is discontinuous at $x = \pm v$, we can remove its discontinuity and
evaluate its values  by applying L'Hospital's rule, which are as follows :

\begin{eqnarray*}
J(x=+v)=\frac{u\{2 u-v(v-2)(v-1)\}}{(v-2)\{2 u^2+(v-1)(v^3+uv)\}}-\frac{uv(v-1)(4v^3-3v^2+uv)\{u+(v-1)v\}}{(v-2)\{2 u^2+(v-1)(v^3+u v)\}^2}
\end{eqnarray*}

and

\begin{eqnarray*}
J(x=-v)=-\frac{u\{2 u+v(v+2)(v+1)\}}{(v+2)\{2u^2+(v+1)(v^3+uv)\}}-\frac{uv(v+1)(4v^3+3v^2+uv)\{u+(v+1)v\}}{(v+2)\{2 u^2+(v+1)(v^3+u v)\}^2}
\end{eqnarray*}

Further, by differentiating $J(x)$ and subsequently applying L'Hospital's rule once again, it can be
shown that $J(x)$ has maxima at $x = + v$ and $x = - v$ and all other values of $J(x)$ within the
domain $x = +\infty$ to $x = -\infty$ are less than this maxima. These maxima values are tabulated
in Table 1. We may note that, these values are much smaller than 1 and actually become still smaller
compared to 1, when either $x >> v$ or the field is weaker or both. So we may safely assume that,
$J(x) << 1 $ for all $x$,  for all practical purposes. In Table 1, we find for \emph{Sun} the difference in
$J(x)$ values for $x = +v$ and $x = -v$, occurs only at 16th place after the decimal. For other
objects we report $J(x)$ values up to 5th place after the decimal.

 Therefore, from Eqns.(\ref{E29}) and (\ref{E30}) one can write:

\begin{eqnarray}\label{E32}
\triangle \psi =& 2 D_{k} \int^{\infty}_{v} \frac{ dx}{x\sqrt{n_{0}^{2}(x) x^{2}-D_{0}^2}}[1 -\frac{1}{2} J(x)+\frac{3}{8}J^{2}(x)-\frac{5}{16}J^{3}(x)+\frac{35}{128}J^{4}(x)-\frac{63}{254}J^{5}(x)+........] - \pi\nonumber\\
=&2[I_0+I_1+I_2+I_3+............]  - \pi
\end{eqnarray}

where, we have introduced additional notations:

\begin{table}
\begin{center}
\footnotesize
\caption{\label{tab:table3}%
 Values of $S_{x}$ and $J(x)$  of different Gravitating Objects  for $x=-v$ and $x=+v$ }
\begin{tabular}{l|c|c|c|c}
  \hline
  & & & &\\
\textrm{Name of Objects}&
\textrm{$S_{x}$ $(x=-v)$} &
\textrm{$S_{x}$ $(x=+v)$} &
\textrm{$J$ $(x=-v)$} &
\textrm{$J$ $(x=+v)$}\\

  \hline

           $\textbf{SUN}$ &
                     0.10241529958672695e-10 &
                     0.10241529958672697e-10 &
                     0.307243e-10&
                     0.307248e-10\\
 & & & &\\
  \hline

$\textbf{PSR B}$ $\textbf{1919+21}$ &
                     0.02720094 &
                     0.02749950  &
                     0.05714 &
                     0.12267 \\
\citep{Hew68} & &   & &\\

\hline

$\textbf{PSR J}$ $\textbf{1748-2446}$ $\textbf{ad}$  &
                     0.02503882  &
                     0.02528190  &
                     0.05342  &
                     0.11053 \\
\citep{Hes06} & &   & &\\

\hline

$\textbf{PSR B}$ $\textbf{1937+21}$ &
                     0.02234780  &
                     0.02254167  &
                     0.04809  &
                     0.09890 \\
\citep{Ash83} & &   & &\\

\hline

$\textbf{PSR J}$ $\textbf{1909-3744}$&
                     0.01235494  &
                     0.01239606  &
                     0.02940  &
                     0.04735 \\
\citep{Jac03} & &   & &\\

\hline

$\textbf{PSR}$ $\textbf{1855+09}$ &
                     0.00630369  &
                     0.00631039  &
                     0.01633  &
                     0.02181 \\
 & & & &\\
\hline

$\textbf{PSRJ}$ $\textbf{0737-3039 A}$&
                     0.00146678  &
                     0.00146691  &
                     0.00417  &
                     0.00461 \\
\citep{Lyn04} &   & & &\\

\hline

$\textbf{PSR}$ $\textbf{0531+21}$ &
                     0.00100644  &
                     0.00100648  &
                     0.00289  &
                     0.00313 \\
  & & & &\\
\hline

$\textbf{PSR B}$ $\textbf{1534+12}$ &
                     0.00087751  &
                     0.00087755  &
                     0.00252  &
                     0.00273  \\
\citep{Sta98} & &   & &\\
\hline
\end{tabular}
\end{center}
\end{table}

\begin{mathletters}
\label{eq:whole}
\begin{equation}\label{E33a}
I_0= D_k \int ^{\infty}_{v}\frac{ dx}{x\sqrt{n_{0}^{2}(x)
x^{2}-D_{0}^2}}\label{subeq:1}
\end{equation}

\begin{equation}\label{E33b}
I_1= D_k \int ^{\infty}_{v}\frac{ dx}{x\sqrt{n_{0}^{2}(x)
x^{2}-D_{0}^2}}(-1/2 J(x))\label{subeq:2}
\end{equation}

\begin{equation}\label{E33c}
I_2= D_k \int ^{\infty}_{v}\frac{ dx}{x\sqrt{n_{0}^{2}(x)
x^{2}-D_{0}^2}}(3/8 J^2(x))\label{subeq:3}
\end{equation}
\end{mathletters}

and so on.

Again we follow a procedure same as what was
followed by Sen\citep{Sen10} to evaluate a similar integral.

Thus

\begin{eqnarray}\label{E34}
I_0=&& D_{k}\int ^{\infty}_{v}\frac{ dx}{x\sqrt{(\frac{x}{x-1})^{2} x^{2}-D_{0}^2}}\nonumber\\
=&&  D_{k}\int ^{\infty}_{v}\frac{(x-1) dx}{ x^{3}\sqrt{1-\frac{D_{0}^2 (x-1)^{2}}{x^{4}}}}\nonumber\\
=&& D_{k} \int ^{\infty}_{v}\frac{(x-2) dx}{ x^{3}\sqrt{1-\frac{D_{0}^2 (x-1)^{2}}{x^{4}}}}+ D_{k}\int ^{\infty}_{v}\frac{ dx}
{ x^{3}\sqrt{1-\frac{D_{0}^2 (x-1)^{2}}{x^{4}}}}\nonumber\\
=&& I_{01}+I_{02} ( say)
\end{eqnarray}

The first part in above (viz $I_{01}$) contains an integral,
which  has been already evaluated earlier by Sen \citep{Sen10}.
Now we introduce a new variable $y=D_{0} x^{-2}(x-1)$,
so that $dy=-D_{0}x^{-3}(x-2)dx$. Therefore,
the  limits of integration change as $y=0$ and

\begin{eqnarray*}
y=&&D_{0} \cdot v^{-2}(v-1)\nonumber\\
 =&&{n_{0}(v)\cdot v}\cdot  v^{-2}(v-1)\nonumber\\
 =&&{\frac{v}{v-1} \cdot v}\cdot v^{-2}(v-1)\nonumber\\
 =&&1
 \end{eqnarray*}

Thus,

\begin{eqnarray}\label{E35}
I_{01}=&&D_{k}\int ^{\infty}_{v}\frac{(x-2) dx}{ x^{3}\sqrt{1-\frac{D_{0}^2 (x-1)^{2}}{x^{4}}}}\nonumber\\
=&&\frac{D_{k}}{D_{0}}\int ^{1}_{0}\frac {dy}{\sqrt{1-y^2}}\nonumber\\
=&&\frac{D_{k}}{D_{0}}\cdot \frac{\pi}{2}
\end{eqnarray}

In the expression for $I_{02}$, if we change the variable as
$z=\frac{1}{x}$, then the limits of this integration changes
from $z=\frac{1}{v}=a (say)$ to $z=0$. A a result, we get a
form of integral as obtained by Sen in his Eqn.(\ref{E17}) \citep{Sen10}.
We can therefore, write:

\begin{eqnarray}\label{E36}
I_{02}=&&D_{k} \int ^{\infty}_{v}\frac{ dx}
{ x^{3}\sqrt{1-\frac{D_{0}^2 (x-1)^{2}}{x^{4}}}}\nonumber\\
=&& D_{k}\int ^{\infty}_{v}\frac{  dx}{ \sqrt{x^{6}-D_{0}^2x^{2}(x-1)^{2}}}\nonumber\\
=&& D_{k}\int ^{a}_{0}\frac{ z dz}{ \sqrt{1-D_{0}^2z^{2}(1-z)^{2}}}
\end{eqnarray}

This can be evaluated in terms of Elliptical function
as expressed by Eqn.(\ref{E18}) of Sen\citep{Sen10}. And finally for
a given value of $a$, its numerical value can be obtained.

Now to evaluate the value of $J(x)$ we evaluate the
value of $((1+2S_{x})^{-1}-1)$ first as follows:

\begin{eqnarray}\label{E37}
(1+2S_{x})^{-1}-1=&&\frac{1}{1+2S_{x}}-1\nonumber\\
=&&-\frac{\frac{2u(xv+u-v)}{(x-1)(x^{3}-uv)}}{1+\frac{2u(xv+u-v)}{(x-1)(x^{3}-uv)}} \nonumber\\
=&&-\frac{2u(xv+u-v)}{(x-1)(x^{3}+uv)+2u^{2}}
\end{eqnarray}

Now in Eqn. (\ref{E31}) above, we substitute the value of
$((1+2S_{x})^{-1}-1)$ (from Eqn. (\ref{E37}))
and $n_0(x)=x/(x-1)$. As a result,
we can write the following expressions for $J(x)$:

\begin{eqnarray}\label{E38}
J(x)=&&\frac{n_{0}^{2}(x) x^{2} \{(1+ 2S_{x})^{-1}  -  1\}+D_{0}^2-D_{k}^2 }{n_{0}^{2}(x) x^{2}-D_{0}^2}\nonumber\\
=&&\frac{n_{0}^{2}(x) x^{2} \{(1+ 2S_{x})^{-1}  -  1\}+D_{0}^2-D_{0}^2 (1+ 2S_{v})^{-1} }{n_{0}^{2}(x) x^{2}-D_{0}^2}\nonumber\\
=&&\frac{  \frac{x^{4}}{(x-1)^2} \{-\frac{2u(xv+u-v)}{(x-1)(x^{3}+uv)+2u^{2}} \}+D_{0}^2\{\frac{2u(v^2+u-v)}{(v-1)(v^3+uv)+2u^2}\} }{\frac{x^{4}}{(x-1)^2}-D_{0}^2}\nonumber\\
=&&\frac{ -  x^{4} \{\frac{2u(xv+u-v)}{(x-1)(x^{3}+uv)+2u^{2}} \}+D_{0}^2 (x-1)^2\{\frac{2u(v^2+u-v)}{(v-1)(v^3+uv)+2u^2}\} }{ x^{4}-D_{0}^2 (x-1)^2}
  \end{eqnarray}

Further, substituting the value of $n_{0}(x)=\frac{x}{x-1}$ and $J(x)$
from Eqn. (\ref{E38}), the integral $I_{1}$ becomes

\begin{eqnarray}\label{E39}
I_{1}=&&- \frac{1}{2} D_{k}\int^{\infty}_{v} \frac{ J(x)}{x\sqrt{n_{0}^{2} x^{2}-D_{0}^2}}dx\nonumber\\
=&& - \frac{1}{2} D_{k}\int^{\infty}_{v} \frac{ 1}{x\sqrt{  \frac{x^{4}}{(x-1)^{2}}-D_{0}^2}}
[\frac{ -  x^{4} \{\frac{2u(xv+u-v)}{(x-1)(x^{3}+uv)+2u^{2}} \}+D_{0}^2 (x-1)^2\{\frac{2u(v^2+u-v)}{(v-1)(v^3+uv)+2u^2}\} }{ x^{4}-D_{0}^2 (x-1)^2}]dx\nonumber\\
=&&-  \frac{1}{2} D_{k}\int^{\infty}_{v} \frac{(x- 1)}{\sqrt{   x^{6}-D_{0}^2 x^{2}(x-1)^{2}}}
[\frac{ -  x^{4} \{\frac{2u(xv+u-v)}{(x-1)(x^{3}+uv)+2u^{2}} \}+D_{0}^2 (x-1)^2\{\frac{2u(v^2+u-v)}{(v-1)(v^3+uv)+2u^2}\} }{ x^{4}-D_{0}^2 (x-1)^2}]dx \nonumber\\
\end{eqnarray}

Again applying the change of variable as $z=\frac{1}{x}$ (like it was done for $I_{02}$)
we may write the integral $I_1$  as

\begin{eqnarray}\label{E40}
I_{1}=&&-  \frac{1}{2}D_{k}\int^{a}_{0} \frac{( 1 -z)}{\sqrt{   1 -D_{0}^2 z^{2}(1 -z)^{2}}}
[\frac{ -    \{\frac{2u z^{3} (v+(u-v)z)}{(1-z)(1+uv z^{3})+2u^{2} z^{4}} \}+D_{0}^2 z^{2} (1-z)^2\{\frac{2u(v^2+u-v)}{(v-1)(v^3+uv)+2u^2}\} }{ 1-D_{0}^2 z^{2} (1-z)^2}]dz \nonumber\\
\end{eqnarray}

Similarly, $I_{2}$ , $I_{3}$ etc. can be written as

\begin{eqnarray}\label{E41}
I_{2}=&&  \frac{3}{8}D_{k}\int^{a}_{0} \frac{( 1 -z)}{\sqrt{   1 -D_{0}^2 z^{2}(1 -z)^{2}}}
[\frac{ -    \{\frac{2u z^{3} (v+(u-v)z)}{(1-z)(1+uv z^{3})+2u^{2} z^{4}} \}+D_{0}^2 z^{2} (1-z)^2\{\frac{2u(v^2+u-v)}{(v-1)(v^3+uv)+2u^2}\} }{ 1-D_{0}^2 z^{2} (1-z)^2}]^{2}dz \nonumber\\
\end{eqnarray}

\begin{eqnarray}\label{E42}
I_{3}=&&-\frac{5}{16}D_{k}\int^{a}_{0} \frac{( 1 -z)}{\sqrt{   1 -D_{0}^2 z^{2}(1 -z)^{2}}}
[\frac{ -    \{\frac{2u z^{3} (v+(u-v)z)}{(1-z)(1+uv z^{3})+2u^{2} z^{4}} \}+D_{0}^2 z^{2} (1-z)^2\{\frac{2u(v^2+u-v)}{(v-1)(v^3+uv)+2u^2}\} }{ 1-D_{0}^2 z^{2} (1-z)^2}]^{3}dz \nonumber\\
\end{eqnarray}

\begin{eqnarray}\label{E43}
I_{4}=&&\frac{35}{64}D_{k}\int^{a}_{0} \frac{( 1 -z)}{\sqrt{   1 -D_{0}^2 z^{2}(1 -z)^{2}}}
[\frac{ -    \{\frac{2u z^{3} (v+(u-v)z)}{(1-z)(1+uv z^{3})+2u^{2} z^{4}} \}+D_{0}^2 z^{2} (1-z)^2\{\frac{2u(v^2+u-v)}{(v-1)(v^3+uv)+2u^2}\} }{ 1-D_{0}^2 z^{2} (1-z)^2}]^{4}dz \nonumber\\
\end{eqnarray}

\begin{eqnarray}\label{E44}
I_{5}=&&-\frac{63}{256}D_{k}\int^{a}_{0} \frac{( 1 -z)}{\sqrt{   1 -D_{0}^2 z^{2}(1 -z)^{2}}}
[\frac{ -    \{\frac{2u z^{3} (v+(u-v)z)}{(1-z)(1+uv z^{3})+2u^{2} z^{4}} \}+D_{0}^2 z^{2} (1-z)^2\{\frac{2u(v^2+u-v)}{(v-1)(v^3+uv)+2u^2}\} }{ 1-D_{0}^2 z^{2} (1-z)^2}]^{5}dz \nonumber\\
\end{eqnarray}

Thus form Eqn. (\ref{E32}), the expression for deflection of light ray
in Kerr geometry can be expressed as:

\begin{eqnarray}\label{E45}
\triangle \psi
=&&2  [\frac{D_{k}}{D_{0}} \frac{\pi}{2}  + D_{k}\{\int ^{a}_{0}\frac{ z dz}{ \sqrt{1-D_{0}^2z^{2}(1-z)^{2}}}\nonumber\\
 &&- \frac{1}{2} \int^{a}_{0} \frac{( 1 -z)}{\sqrt{   1 -D_{0}^2 z^{2}(1 -z)^{2}}}
(\frac{ -    \{\frac{2u z^{3} (v+(u-v)z)}{(1-z)(1+uv z^{3})+2u^{2} z^{4}} \}+D_{0}^2 z^{2} (1-z)^2\{\frac{2u(v^2+u-v)}{(v-1)(v^3+uv)+2u^2}\} }{ 1-D_{0}^2 z^{2} (1-z)^2})dz  \nonumber\\
&& + \frac{3}{8}\int^{a}_{0} \frac{( 1 -z)}{\sqrt{   1 -D_{0}^2 z^{2}(1 -z)^{2}}}
(\frac{ -    \{\frac{2u z^{3} (v+(u-v)z)}{(1-z)(1+uv z^{3})+2u^{2} z^{4}} \}+D_{0}^2 z^{2} (1-z)^2\{\frac{2u(v^2+u-v)}{(v-1)(v^3+uv)+2u^2}\} }{ 1-D_{0}^2 z^{2} (1-z)^2})^{2}dz \nonumber\\
&& -\frac{5}{16}\int^{a}_{0} \frac{( 1 -z)}{\sqrt{   1 -D_{0}^2 z^{2}(1 -z)^{2}}}
(\frac{ -    \{\frac{2u z^{3} (v+(u-v)z)}{(1-z)(1+uv z^{3})+2u^{2} z^{4}} \}+D_{0}^2 z^{2} (1-z)^2\{\frac{2u(v^2+u-v)}{(v-1)(v^3+uv)+2u^2}\} }{ 1-D_{0}^2 z^{2} (1-z)^2})^{3}dz \nonumber\\
&& + \frac{35}{64} \int^{a}_{0} \frac{( 1 -z)}{\sqrt{   1 -D_{0}^2 z^{2}(1 -z)^{2}}}
[\frac{ -    \{\frac{2u z^{3} (v+(u-v)z)}{(1-z)(1+uv z^{3})+2u^{2} z^{4}} \}+D_{0}^2 z^{2} (1-z)^2\{\frac{2u(v^2+u-v)}{(v-1)(v^3+uv)+2u^2}\} }{ 1-D_{0}^2 z^{2} (1-z)^2}]^{4}dz\nonumber\\
&&-\frac{63}{256} \int^{a}_{0} \frac{( 1 -z)}{\sqrt{   1 -D_{0}^2 z^{2}(1 -z)^{2}}}
[\frac{ -    \{\frac{2u z^{3} (v+(u-v)z)}{(1-z)(1+uv z^{3})+2u^{2} z^{4}} \}+D_{0}^2 z^{2} (1-z)^2\{\frac{2u(v^2+u-v)}{(v-1)(v^3+uv)+2u^2}\} }{ 1-D_{0}^2 z^{2} (1-z)^2}]^{5}dz\nonumber\\
&&+.........\} ] - \pi\nonumber\\
or, \triangle \psi
=&& \frac{D_{k}}{D_{0}} \pi  + 2D_{k}[\int ^{a}_{0}\frac{ z dz}{ \sqrt{1-D_{0}^2z^{2}(1-z)^{2}}}\nonumber\\
 &&- \frac{1}{2}\int^{a}_{0} \frac{( 1 -z)}{\sqrt{   1 -D_{0}^2 z^{2}(1 -z)^{2}}}
(\frac{ -    \{\frac{2u z^{3} (v+(u-v)z)}{(1-z)(1+uv z^{3})+2u^{2} z^{4}} \}+D_{0}^2 z^{2} (1-z)^2\{\frac{2u(v^2+u-v)}{(v-1)(v^3+uv)+2u^2}\} }{ 1-D_{0}^2 z^{2} (1-z)^2})dz  \nonumber\\
&& + \frac{3}{8}\int^{a}_{0} \frac{( 1 -z)}{\sqrt{   1 -D_{0}^2 z^{2}(1 -z)^{2}}}
(\frac{ -    \{\frac{2u z^{3} (v+(u-v)z)}{(1-z)(1+uv z^{3})+2u^{2} z^{4}} \}+D_{0}^2 z^{2} (1-z)^2\{\frac{2u(v^2+u-v)}{(v-1)(v^3+uv)+2u^2}\} }{ 1-D_{0}^2 z^{2} (1-z)^2})^{2}dz \nonumber\\
&& -\frac{5}{16}\int^{a}_{0} \frac{( 1 -z)}{\sqrt{   1 -D_{0}^2 z^{2}(1 -z)^{2}}}
(\frac{ -    \{\frac{2u z^{3} (v+(u-v)z)}{(1-z)(1+uv z^{3})+2u^{2} z^{4}} \}+D_{0}^2 z^{2} (1-z)^2\{\frac{2u(v^2+u-v)}{(v-1)(v^3+uv)+2u^2}\} }{ 1-D_{0}^2 z^{2} (1-z)^2})^{3}dz\nonumber\\
&& + \frac{35}{64} \int^{a}_{0} \frac{( 1 -z)}{\sqrt{   1 -D_{0}^2 z^{2}(1 -z)^{2}}}
[\frac{ -    \{\frac{2u z^{3} (v+(u-v)z)}{(1-z)(1+uv z^{3})+2u^{2} z^{4}} \}+D_{0}^2 z^{2} (1-z)^2\{\frac{2u(v^2+u-v)}{(v-1)(v^3+uv)+2u^2}\} }{ 1-D_{0}^2 z^{2} (1-z)^2}]^{4}dz\nonumber\\
&&-\frac{63}{256} \int^{a}_{0} \frac{( 1 -z)}{\sqrt{   1 -D_{0}^2 z^{2}(1 -z)^{2}}}
[\frac{ -    \{\frac{2u z^{3} (v+(u-v)z)}{(1-z)(1+uv z^{3})+2u^{2} z^{4}} \}+D_{0}^2 z^{2} (1-z)^2\{\frac{2u(v^2+u-v)}{(v-1)(v^3+uv)+2u^2}\} }{ 1-D_{0}^2 z^{2} (1-z)^2}]^{5}dz\nonumber\\
&&+......... ]- \pi\nonumber\\
\end{eqnarray}

The above expression has been obtained for
gravitational deflection in the equatorial plane
of a rotating body considering the Kerr line element.

\section{\label{4}Calculation of numerical values for deflection for some Gravitational objects}

Considering $Sun$ as a rotating body, we can calculate
 the deflection angle for pro-grade $(b=-ve)$ and
 retro-grade $(b=+ve)$ direction of the light ray
 with respect to the gravitating body. From Eqn.(\ref{E45})
after actual numerical calculations, it was found that
for pro-grade direction, the deflection will be greater
and for retro-grade direction, the deflection will
be smaller than that of the Schwarzschild geometry.
This has been also confirmed by Iyer
$\textrm{$et$ $ al.$}$ in their previous work \citep{Iye09}
and also by Alsing \citep{Als98} in a similar work.
A more recent work by Werner \citep{Wer12}, also confirms this
phenomena where the author has considered terms only
up to first order in $\alpha$ ( rotation parameter).

For a Sun grazing ray, we can calculate the
deflection of a light ray using the Eqn.(\ref{E45})
obtained in this present work. Here, we may
consider the closest distance of approach is equal
to the solar radius as $r_{\bigodot}=6.955\times10^5$ km,
solar mass as $M_{\bigodot}=1.99\times10^{30}$ kg and
solar time period as $T=28$ days.

In addition to $Sun$, we shall also consider some
millisecond pulsars, to show the effect of rotation
on light deflections, as we know the pulsars are
fast rotating objects. Nunez $\textrm{$et$ $ al.$}$ \citep{Nun10}
have calculated the red-shift and  preferred  radius
for some fast millisecond pulsars. We derive some
of the input parameters ( like $b$, $\alpha$, $r_g$ etc.)
from their published work and perform some sample calculations.
We note that, the calculations performed here on
$Sun$ and these pulsars are for demonstrative purpose
only. The aim is to show that, like the {\it Null Geodesic}
method, the {\it Material Medium approach}
can  predict equally well the
gravitational deflection of the light ray in Kerr field.

\begin{deluxetable}{l|l|l|l|l|l|l|l|l}
\tabletypesize{\scriptsize}
\rotate
\tablecaption{Refractive index due to different $Gravitational$ $Objects$.}\label{tbl-1}
\tablewidth{0pt}
\tablehead{
\textrm{Name of the}&
\textrm{Schwarzs} &
\textrm{Rotation} &
\textrm{Physical} &
\textrm{$u$} &
\textrm{$v$} &
\textrm{Refractive}&
\textrm{Refractive}&
\textrm{Refractive}\\
\textrm{Gravitational}&
\textrm{child} &
\textrm{Parameter} &
 \textrm{Radius$R$} &
 \textrm{$(\alpha/r_g)$} &
 \textrm{$(b/r_g)$} &
 \textrm{ index n} &
 \textrm{ index n} &
 \textrm{ index n}\\
 \textrm{Objects}&  Radius &  & or & & &  &  (Pro-grade) & (Retro-grade)\\
 &  $r_g$ $(Km)$ & $\alpha$ $ (Km) $ & $b$ $(Km)$ & & &  $(\alpha=0)$ &$(b=-ve)$ & $(b=+ve)$
}
\startdata
$SUN$ & 2.9554 & 1.6762 & 6.955e+05  &  0.5671  & 235331.2911 & 1.00000424934 & 1.00000424935 & 1.00000424933\\
 &   &   &  &  &  &  &   & \\
 \hline

 $PSRB$ $1919+21$& 4.1375 & 2.5075 & 2.000e+01   &  0.6060  &      4.8337 &1.26084097201 & 1.29285456990 & 1.22753613865\\
\citep{Hew68}& &   &  &  &  &  &   & \\

  \hline

 $PSRJ$ $1748-2446$ & 3.9898 & 2.4261 & 2.010e+01  &  0.6080  &     5.0378  & 1.24765683468 & 1.27698068171 & 1.21726163358\\
\citep{Hes06} & &   &  &  &  &    & &\\

 \hline

$PSRB$ $1937+21$  & 3.9898 & 2.1970 & 2.020e+01 &  0.5506 &  5.0629 & 1.24612905064 & 1.27244504736 & 1.21895467231\\
\citep{Ash83} & &   &  &  &  &  &   & \\

 \hline

$PSR J$ $1909-3744$&  4.2498  & 2.7486 & 3.110e+01 &  0.6467 & 7.3178  & 1.15828146469  & 1.17215555862 & 1.14418491095\\
\citep{Jac03} & &   &  &  &  &  &    &\\

 \hline

$PSR$ $1855+09$ &  3.9898  & 3.4339 & 4.690e+01 & 0.8606 &    11.7549 & 1.09298024705 & 1.09976248166 & 1.08614772138\\
& &   &  &  &  &  &    &\\

 \hline

 $PSRJ$ $ 0737-3039$ & 3.9602 & 6.5918 & 1.336e+02  &  1.6645 & 33.7352   & 1.03054808987 &  1.03205416015 & 1.02903968046 \\
\citep{Lyn04} & &   &  &  &  &    & &\\

 \hline

 $PSR$ $0531+21$& 3.9898 & 6.8042 & 1.644e+02 & 1.7054 & 41.2050  & 1.02487248899 & 1.02590137268 & 1.02384251987 \\
& &   &  &  &  &  &    &\\

 \hline

 $PSR B$ $1534+12$ &  3.9602 & 6.0879 & 1.659e+02 & 1.5372 &  41.8913 & 1.02445506278 & 1.02535206812  & 1.02355723286\\
\citep{Sta98} &   &   &  &  &  &     & &\\
\enddata
\end{deluxetable}

In table 2, we have shown the refractive
index along with other parameters
for different gravitating body, due to pro-grade and
retro-grade orbits of the light ray.
The refractive index values
have been expressed with 11 places after the decimal.
We can note for $Sun$ the refractive index value
changes at 11th place after the decimal as we change
from  pro-grade to retro-grade orbit of light ray.
However, for fast rotating pulsars these refractive
index values vary at the second place after the decimal
between pro-grade and retrograde orbits ( cf. Table 2).
From this table it is totally clear that, the refractive
index is greater for pro-grade direction and smaller for
retro-grade direction as compared to Schwarzschild
one. Thus as compared to the Schwarzschild case,
the deflection angle for light ray should be also
greater for pro-grade and smaller for retro-grade
orbits of light ray.

In table 3, we have calculated the deflection of
a light ray (in arc-sec) due to \textrm{\emph{Sun}} and other pulsars
considering the impact parameter as the physical
radius of the gravitating body and considering up to
fifth order term i.e. $I_5$ in the Eqns. (\ref{E32}) and (\ref{E45}).
We further, note that $Sun$ is a
slow rotating object and pulsars are fast rotating. To
calculate the deflection angle of these rotating
gravitational objects, we used numerical integration
by Simpson's one third rule. Here it is also
shown that, the value of deflection angle continuously
decreases as we calculate higher order terms. In
this table the deflection angle values due to
the gravitational body are calculated considering
no rotation ($\alpha=0$) of the body, pro-grade ($b=-ve$) and
retro-grade ($b=+ve$) orbits of light ray. In case of no
rotation ( $\alpha=0$ ), the first column ( representing
$(2I_0-\pi)$) will be zero as
there is no difference between $D_{k}$ and $D_{0}$.
Also the values of $I_1$, $I_2$ , $I_3$ etc.  will be zero,
when there is no rotation ( i.e. $\alpha= 0$) as can be seen
in Table 3 and equally confirmed through Eqn (\ref{E45}).
The light deflection values obtained for Sun, by
material medium method here for pro-grade and
retro-grade orbits are  in good agreement
with those other similar calculations done using
null geodesic method [Iyer \& Hansen\citep{Iye09}] .

\begin{deluxetable}{l|c|c|c|c|c|c|c|c}
\tabletypesize{\scriptsize}
\rotate
\tablecaption{\label{tab:table2}
Deflection of a light ray due to   $Sun$ and some $Pulsars$}
\tablewidth{0pt}
\tablehead{
 \textrm{Name } &
\textrm{$2I_{01}-\pi$} &
\textrm{$2I_{02}$} &
\textrm{$2I_1$} &
\textrm{$2I_2$} &
\textrm{$2I_3$} &
\textrm{$2I_4$} &
\textrm{$2I_5$} &
\textrm{Total Deflection}\\
\textrm{of} &\textrm{ (a)} & \textrm{ (b)} & \textrm{ (c)} &\textrm{ (d)} & \textrm{(e)} &
\textrm{ (f)} & \textrm{ (g)} & \textrm{(a+b+...+f+g)}\\
 \textrm{Objects} &  \textrm{(arc-sec)}  & \textrm{(arc-sec)} & \textrm{(arc-sec)}  & \textrm{(arc-sec)}  & \textrm{(arc-sec)} & \textrm{(arc-sec)}  & \textrm{(arc-sec)} & \textrm{$\triangle \phi $(arc-sec)}}
\startdata
  $\textbf{Sun}$ & 0.00000 &    1.752008089976  &   &   &  &  &  &  1.752008089976\\
 $\alpha=0$ &  &  & & & & & &\\
 \hline
$Pro$ & 0.66364e-05 &  1.752008089994 & -0.18122e-05 & 0.10665e-16 & -0.40393e-28 & 0.12069e-38 & -0.50226e-50 &  1.752012914194 \\
($b$=-ve)   &  & & & & & & &\\
\hline
$Retro$ & -0.66365e-05 &   1.752008089959   & -0.18122e-05  & 0.10665e-16 & -0.76563e-28 & 0.12069e-38 & -0.50226e-50  & 1.751999641259 \\
 ($b$=+ve) &  &  & & & & & &\\
 \hline
\hline
 $\textbf{B1919}$   & 0.00000 & 119735.4072 &   &   &  &  &  & 119735.4072\\
$\alpha=0$ &  &  & & & & & &\\
 \hline
$Pro$ & 16453.1545 & 122775.5695 & -10484.7664 & 307.7444 & -5.3394 & 0.8887 & -0.1876e-01  &
  129047.2325 \\
($b$=-ve)   &  &  & & & & & &\\
\hline
$Retro$ & -17116.7756   & 116572.6231 & -9955.0483 & 292.1964 & -10.5747 & 0.8438 & -0.1781e-01 &
 89783.2468  \\
($b$=+ve)  &  &  & & & & & &\\
 \hline
 \hline
 $\textbf{J1748}$ & 0.00000 & 112974.7176   &   &   &  &  &  &  112974.7176\\
  $\alpha=0$  &  &  & & & & & &\\
 \hline
$Pro$ & 15230.0314 & 115629.9776 & -9290.5744 &  243.6810 & -3.7975 & 0.5612 & -0.1057e-01 &
  121809.8687\\
($b$=-ve)  &  &  & & & & & &\\
\hline
$Retro$ &  -15786.4644 & 110222.4469 & -8856.0931 & 232.2850 & -7.5082 & 0.5350 & -0.1008e-01 &
 85805.1911  \\
 ($b$=+ve) &  &  & & & & & &\\
 \hline
\hline
 $\textbf{B1937}$ & 0.00000 & 112197.2094   &   &   &  &  &  &  112197.2094\\
  $\alpha=0$  &  &  & & & & & &\\
 \hline
$Pro$ & 13684.5905 & 114566.6120 &  -8286.1164 & 194.5455 & -2.7143 & 0.3591 & -0.6060e-02  &
  120157.2703\\
($b$=-ve)  &  &  & & & & & &\\
\hline
$Retro$ &  -14130.9579 & 109750.5211 & -8751.2927 & 226.4006 & -7.2176 & 0.5072 & -0.9428e-02 &
 87087.9512  \\
($b$=+ve) &  &  & & & & & &\\
 \hline
 \hline
$\textbf{J1909}$ & 0.00000 & 69386.1605 &   &   &  &  &  &  69386.1605\\
   $\alpha=0$ &  &  & & & & & &\\
 \hline
$Pro$ &  7761.8550 & 70217.2798 & -3538.9772 & 37.3630 & -0.2419 & 0.1382e-01 & -0.1043e-03 &
  74477.2924\\
($b$=-ve)  &  &  & & & & & &\\
\hline
$Retro$ &   -7886.3101 & 68541.7148 & -3454.5281 &  36.4714 & -0.4728 & 0.1349e-01 & -0.1008e-03 &
  57236.8885 \\
($b$=+ve)  &  &  & & & & & &\\
 \hline
 \hline
$\textbf{1855}$  & 0.00000 & 39691.1980 &   &   &  &  &  &  39691.1980\\
  $\alpha=0$  &  &  & & & & & &\\
 \hline
$Pro$ &  4021.0132 & 39937.4925 & -1486.8124 &  6.8396 & -0.1971e-01 & 0.4775e-03 & -0.1563e-05 &
  42478.5136 \\
($b$=-ve)  &  &  & & & & & &\\
\hline
$Retro$ &  -4050.8295 & 39443.0772 & -1468.4061 &  6.7549 & -0.3805e-01 & 0.4715e-03 & -0.1543e-05 &
 33930.5589   \\
($b$=+ve) &  &  & & & & & &\\
 \hline
\hline
 $\textbf{J0737}$ & 0.00000 & 12734.8996    &   &   &  &  &  &  12734.8996\\
  $\alpha=0$   &  &  & & & & & &\\
 \hline
$Pro$ &  947.0043 &  12753.5108 & -285.9891 & 0.2616 & -0.1527e-03 & 0.7180e-06 & -0.4655e-09 &
 13414.7874\\
($b$=-ve)  &  &  & & & & & &\\
\hline
$Retro$ &  -948.4751 & 12716.2596 & -285.1537 & 0.2608 & -0.2915e-03 & 0.7159e-06 & -0.4642e-09 &
 11482.8913 \\
($b$=+ve) &  &  & & & & & &\\
 \hline
 \hline
$\textbf{0531}$  & 0.00000 & 10347.3213 &   &   &  &  &  &  10347.3213\\
  $\alpha=0$ &  &  & & & & & &\\
 \hline
$Pro$ & 650.5361 & 10357.7091 & -192.8309 & 0.1193 & -0.4718e-04 & 0.1496e-06 & -0.6561e-010 &
 10815.5335 \\
($b$=-ve)  &  &  & & & & & &\\
\hline
$Retro$ &  -651.2224 & 10336.9225 & -192.4439 &  0.1190  & -0.8998e-04 & 0.1493e-06 & -0.6548e-010 &
 9493.3751\\
($b$=+ve) &  &  & & & & & &\\
 \hline
 \hline
$\textbf{1534}$  & 0.00000 & 10172.1154 &   &   &  &  &  &  10172.1154\\
  $\alpha=0$  &  &  & & & & & &\\
 \hline
 $Pro$ & 567.3840 & 10181.0220 & -167.9483 &   0.9054e-01 & -0.3119e-04 & 0.8622e-07 & -0.3293e-010 &
  10580.5482  \\
 ($b$=-ve) &  &  & & & & & &\\
\hline
 $Retro$ &   -567.9056 &  10163.2006 & -167.6543 &  0.9038e-01 & -0.5950e-04 & 0.8607e-07 & -0.3287e-010 &
 9427.7310   \\
($b$=+ve) &  &  & & & & & &\\
\enddata
\end{deluxetable}

\begin{figure}[!htbp]
\centering
\includegraphics{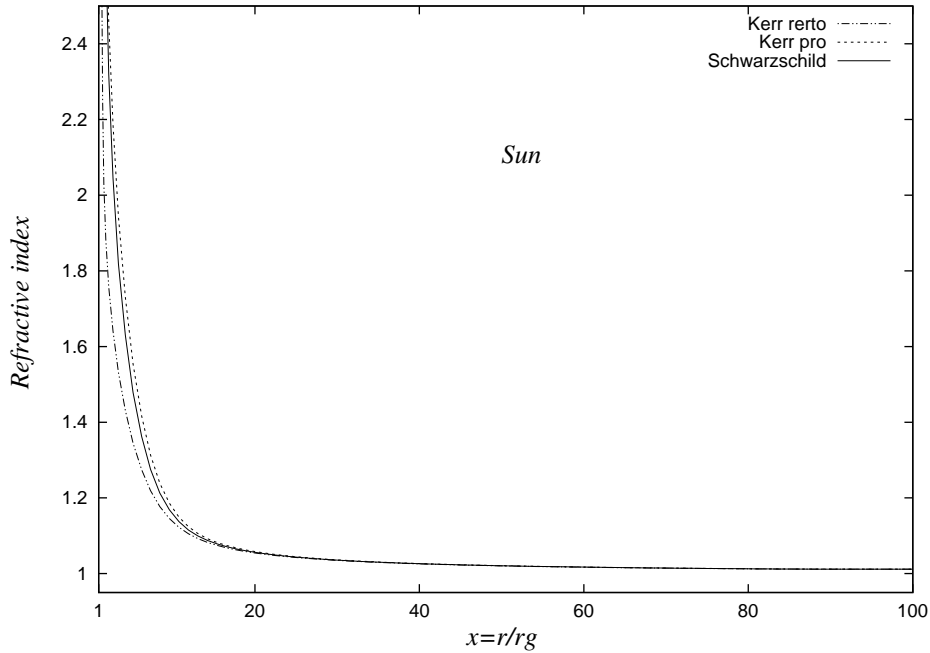}
\caption{Refractive index($n(x,\frac{\pi}{2})$) as a function of $x$($=\frac{r}{r_{g}}$) for Sun.}\label{fig:1}
\end{figure}

\begin{figure}[!htbp]
\centering
\includegraphics{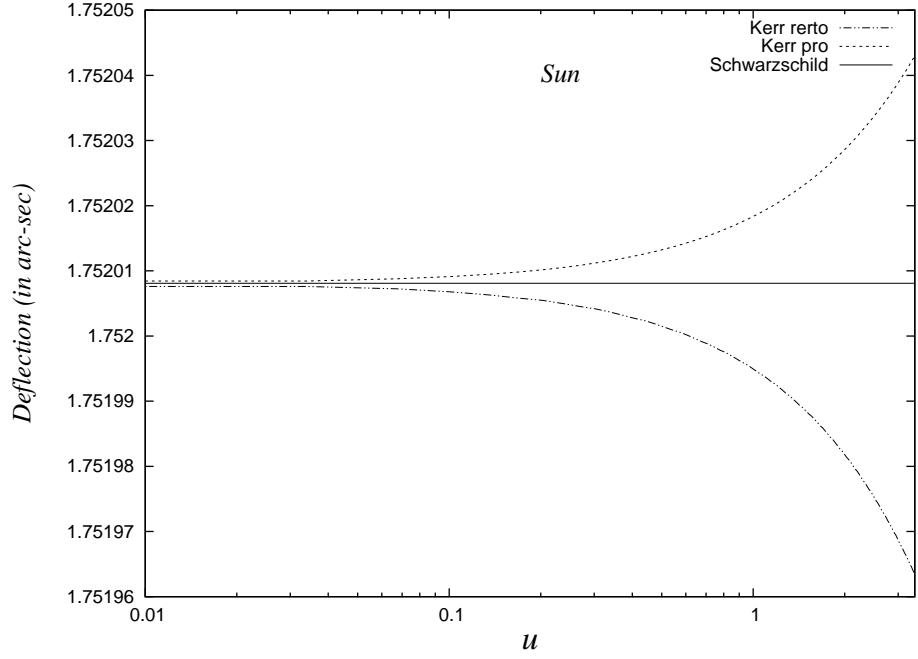}
\caption{Deflection($\triangle \psi$)  as a function of $u$($=\frac{\alpha}{r_{g}}$) for Sun.}\label{fig:2}
\end{figure}

\begin{figure}[!htbp]
\centering
\includegraphics{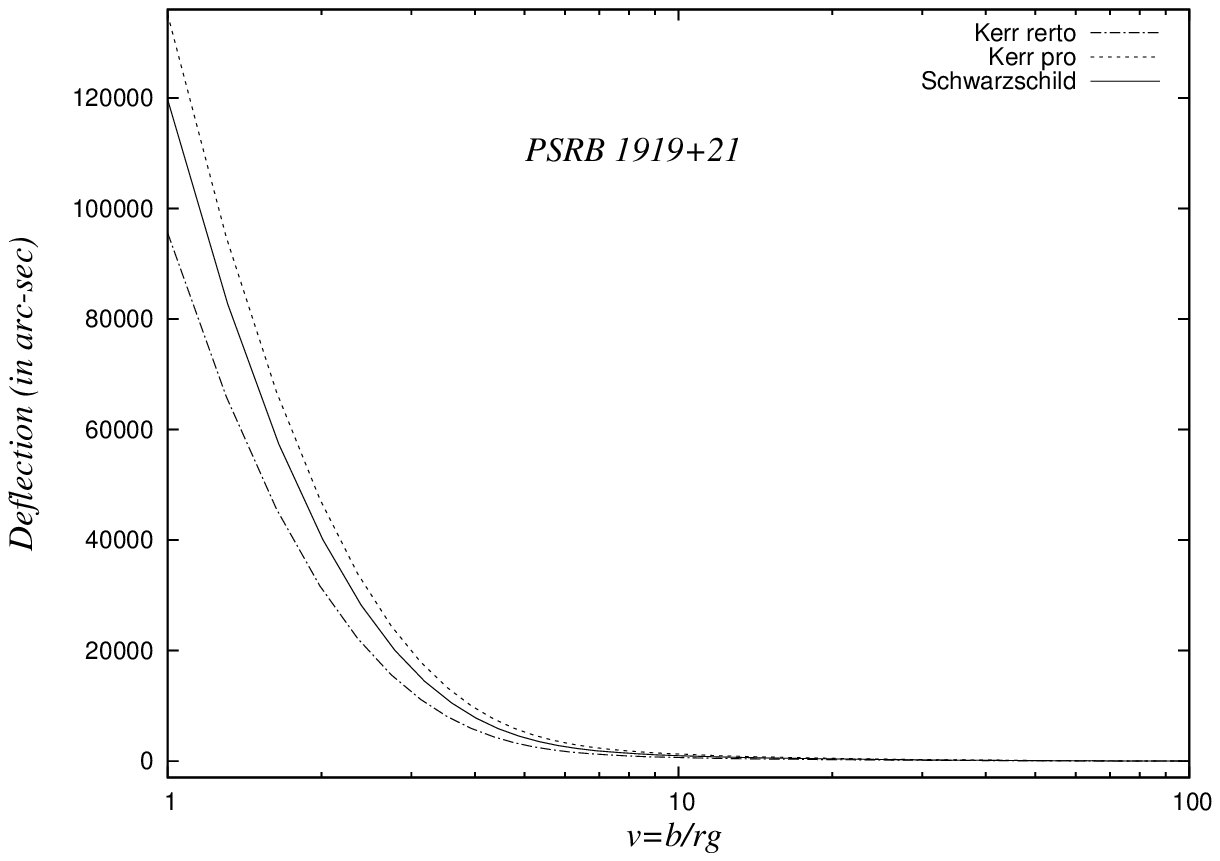}
\caption{Deflection($\triangle \psi$)  as a function of    $v$($=\frac{b}{r_{g}}$) for pulsar $PSRB$ 1919+21.}\label{fig:3}
\end{figure}

In Fig.\ref{fig:1} we have plotted the refractive index as
a function of $x$ considering the gravitational body
to be  $Sun$. If we make this plot for other pulsars,
we get the similar trend in the curves.
 As the value of $x$ i.e. $\frac{r}{r_g}$ increases,
( i.e. as we move towards asymptotically flat space),
these curves merge into each other.
 At $x=1$ i.e. $r=r_{g}$, the value of refractive
 index is infinite for all the gravitational bodies
 and this is physically expected.

In Fig.\ref{fig:2}, we have plotted the deflection
angle as a function of $u ( =\frac{\alpha}{r_{g}})$ for Sun.
Thus, Fig.\ref{fig:2} illustrates the dependence of deflection angle
on the rotation parameter. Here the solid line parallel
to the $u$-axis, indicates the Schwarzschild geometry.

The pulsars are highly compact objects, with their
physical radii being very close to the corresponding
Schwarzschild radii in most of the cases. Thus it
is necessary to consider the light ray to be passing
at various other close distances from the pulsars.
In order to understand the
dependence of deflection on the impact parameter,
we selected one of the pulsars from our list
viz. PSRB 1919+21. In Fig.\ref{fig:3} we have plotted the amount
of deflection as a function of  $v= b/r_g$, under three
cases no-rotation, pro-grade orbit and retro-grade orbit.
From Fig.\ref{fig:3}, it is also  clear that the deflection angle is
greater for pro-grade and smaller for retro-grade
orbit as compared to the Schwarzschild one.
This  phenomena has been also observed by
Iyer $\textrm{$et$ $ al.$}$  \citep{Iye09} as discussed
earlier.

From Fig.\ref{fig:2} and \ref{fig:3}, and also from Table 3, it is clear that
the deviations of light orbit from the Schwarzschild case
for pro-grade and retro-grade orbit are not symmetric.
Such asymmetries of deflection values for pro-grade and
retro-grade orbits have been also reported by previous
authors [Iyer\citep{Iye09}].

\section{\label{5}Conclusions}

We have presented here in detail the calculations
for light deflection angle on the equatorial plane of
a rotating objects (viz Kerr field), by
following {\it material medium approach}.
The following have been observed:

(i) The {\it material medium approach} gives same
values of deflection as compared to that obtained
by other most conventional method of {\it Null geodesic}.
This has been verified by taking {\it Sun} as a test case.
\textit{Further, the cases of some millisecond pulsars were
considered to understand the effect of rotation more objectively
on the deflection angle.}

(ii) For pro-grade orbit of the  deflection
angle is greatest and for retro-grade orbit, the
deflection angle is smallest. The one corresponding to
no-rotation ( Schwarzschild case) lies in between.

(iii) The deviations of pro-grade and retro-grade orbits
from the Schwarzschild deflection angle  are not symmetric.
The deviation  is slightly higher for retro-garde orbit.

\section*{Acknowledgments}

We are thankful to anonymous reviewer of this paper for his/her  useful comments.

\end{document}